\documentstyle[aps,prb,multicol,epsfig]{revtex}

\begin{document}

\title{Disordered periodic systems at the upper critical dimension}

\author{R. Chitra}
\address{Physics Department, Rutgers University, Piscataway, NJ-08854,
USA}
\author{T. Giamarchi}
\address{Laboratoire de Physique des Solides, CNRS-URA 02, UPS Bat. 510,
91405 Orsay
France}
\author{P. Le Doussal}
\address{CNRS-Laboratoire de Physique Theorique de
l'Ecole Normale Superieure,
24 rue Lhomond,75231 Cedex 05, Paris, France.}
\maketitle

\begin{abstract}
The effects of weak point-like disorder on periodic systems at their
upper critical dimension $D_c$ for disorder
are studied. The systems studied
range from simple elastic systems with $D_c=4$ to systems with 
long range interactions with $D_c=2$ and
systems with  $D_c=3$ 
such as the vortex lattice with dispersive elastic constants.
These problems are studied using the Gaussian Variational
method and the Functional Renormalisation Group.
In all the cases studied we find a typical ultra-slow  
$\log \log(x)$ growth of the asymptotic
displacement correlation function, resulting in nearly perfect 
translational order. Consequences for the Bragg glass phase 
of vortex lattices are discussed.
\end{abstract}


\section {Introduction}
The  study of the
physics of elastic systems in the presence of disorder has long been
a subject of great theoretical and experimental interest, with
applications to  systems as diverse as  charge density waves
\cite{gruner_revue_cdw}, Wigner crystals \cite{andrei_wigner_2d},
magnetic bubbles
\cite{seshadri_bubbles_thermal,seshadri_bubbles_long}, etc. .
These studies were given a further impetus with the discovery
of high-Tc superconductors which were found to be of the Type II kind.
The  resolution of the various theoretical challenges posed by these
disordered  vortex lattices
\cite{blatter_vortex_review} is crucial for the
technological applications of high-$T_c$ materials.
Early studies  by Larkin \cite{larkin_70}
of the elastic system with point-like
disorder
indicated that disorder is relevant
and that perfect long range translational order is destroyed below
four dimensions. Since for a periodic system
thermal effects dominate over that of disorder in dimensions $d < 2$,
$d=4$ and $d=2$  are the upper and lower critical
dimension for the relevance of disorder.
However,  the approximate model of
independent random forces acting on each vortex line used in
Ref.~\onlinecite{larkin_70,larkin_ovchinnikov_pinning}
was insufficient to obtain the asymptotic
properties of the disorder dominated phase.
Scaling arguments \cite{nattermann_pinning,villain_cosine_realrg}
suggested, within a
purely elastic description, a slower, logarithmic growth of displacements.
However, general arguments also tended to prove that
disorder \emph{always} favors the presence of
dislocations \cite{villain_cosine_realrg,fisher_vortexglass_long} which
thus invalidates the elastic approximation and 
leads to a destruction of  the translational order. 
Although the precise nature of the disordered
phase was a subject of much debate there was a 
thus a consensus
that  translational order was absent  and that
disorder leads to a glassy state with diverging
barriers and pinning
\cite{fisher_vortexglass_long,feigelman_collective}.
Various experimental features
such as the existence of a first order transition
\cite{charalambous_melting_rc,safar_tricritical_prl},
presence of large  dislocation free regions in decoration experiments
\cite{grier_decoration_manips}, and termination of the 
melting line \cite{safar_tricritical_prl,zeldov_diagphas_bisco}
could not be explained by these theories.

A quantitative theory of the
elastic vortex lattice
with point disorder was recently developed 
\cite{giamarchi_vortex_short,giamarchi_vortex_long}
using both Functional Renormalization Group (FRG) and
Gaussian Variational methods (GVM). These methods were originally  
introduced to study random manifold systems
\cite{fisher_functional_rg,mezard_variational_replica,%
bouchaud_variational_vortex}.
In the case of periodic systems both methods were found to
be in qualitative and quantitative agreement, the
FRG being precise in $d=4-\epsilon$ dimensions and the
variational method being applicable for $2 \leq d < 4$.
These studies \cite{giamarchi_vortex_short,giamarchi_vortex_long}
yielded a description of the vortex lattice at
\emph{all scales} and demonstrated that while disorder produces
algebraic growth of displacements at short length scales,
periodicity changes the growth to logarithmic at large scales
\cite{nattermann_pinning,villain_cosine_realrg,%
giamarchi_vortex_short,giamarchi_vortex_long,korshunov_variational_short}
thereby resulting in a \emph{power law}
decay of translational order. However, due to the abovementioned 
energy arguments 
indicating \cite{villain_cosine_realrg,fisher_vortexglass_long}
that disorder always generate dislocations, it was important 
to determine whether these results, derived in the elastic 
approximation would apply to realistic systems. 
These energy arguments, reexamined in
Ref.~\onlinecite{giamarchi_vortex_long} turned out to be incorrect
and instead it was found that 
dislocations are \emph{unfavourable} for weak disorder in
$d=3$. This led to the prediction of a distinct \emph{thermodynamically}
stable \emph{dislocation free} glass phase called the Bragg glass
because of its \emph{nearly perfect} translational order and perfect
topological order. Furthermore it was proposed 
\cite{giamarchi_vortex_long,giamarchi_diagphas_prb} that the
phase seen experimentally at low fields was the Bragg glass,
which accounted naturally for the first-order transition
and the decoration experiments.
These predictions were confirmed by numerical
\cite{gingras_dislocations_numerics,ryu_diagphas_numerics,vanotterlo_bragg_numerics}
and analytical
\cite{kierfeld_bglass_layered,carpentier_bglass_layered,fisher_bragg_proof,%
ertas_diagphas_bisco,goldschmidt_diagphas_bisco,kierfeld_diagphas_bisco,%
koshelev_diagphas_bisco,vinokur_diagphas_bisco} studies.

These early studies were made using dispersionless elastic constants,
an approximation indeed valid to study the large scale properties of
the system. However, in systems such as high $T_c$ superconductors 
the penetration
depths can be large compared to the lattice constant, resulting in 
the possibility of non-local elastic
constants at intermediate length scales. The effect of such non 
local elasticity
was qualitatively studied in
Ref.~\onlinecite{blatter_vortex_review,giamarchi_vortex_long}
and in more detail recently in
Ref.~\onlinecite{bucheli_frg_rc,wagner_thesis}
where the Larkin length and the pinning force were computed, using a
renormalization group procedure.

In view of the
physical importance and the
remarkable properties of the Bragg glass phase, it is interesting to
compute other physical properties, such as the positional order for
intermediate length scales where the dispersion of the elastic
constants is important. Interestingly, this dispersivity
of the elastic constants pushes the $d=3$ Abrikosov lattice to
its upper critical dimension\cite{bucheli_frg_rc,wagner_thesis}.
We thus address in this paper the more general
question of the effects of disorder right at the upper critical dimension
$D_c$
(for a discussion on
what happens at the lower critical dimension see Ref.
~\onlinecite{giamarchi_vortex_long,ledoussal_rsb_prl,%
giamarchi_book_young,carpentier_ledou_triangco,carpentier_melting_prl}).
Other examples of such systems which mimic the behaviour at $d=4$
are single component elastic systems in  $d=2$ such as charge density waves
where  long range $1/r$ Coulomb interactions
increase the effective dimension of the system to $d=4$ i.e., $D_c=2$.
Other motivations for this problem are theoretical. First,
the logarithmic growth of displacements found for $d < 4$ has
a prefactor proportional to $\epsilon=4-d$.
It is  thus interesting to
study how the displacements grow at  the upper critical
dimension. Second, it also allows us to check and compare the results of
the GVM and the FRG when the disorder is marginal.

The plan of the paper is as follows. Section~\ref{sec:upper}
is devoted to the study of simple isotropic
elastic systems at the upper critical dimension $d=4$
using both the GVM and the FRG.
We compute the correlation functions and the characteristic
length of the system, and compare the two methods.
In Section~\ref{sec:disp} we study  physical systems with
dispersive elastic constants  and anisotropy,
in particular  the $d=3$
vortex lattice in the non local elastic regime.
The conclusions  are presented  in Section~\ref{sec:conc}.

\section{Elastic systems at the upper critical dimension $D_c=4$ }
\label{sec:upper}
\subsection{ The model } \label{sec:model}

We consider the simplest model of an isotropic  elastic system where
deviations  from the equilibrium positions,  characterised
by the variables $u_{\alpha}$, are described in
a coarse-grained manner by the following continuum hamiltonian
\begin{equation}
H= \int d^d x  [\frac{c}2 (\partial u_{\alpha})^2
+    V(x) \rho(x)]
\end{equation}
\noindent
Here $d$ is the spatial dimension and the index $\alpha$ runs over the
number
of components of the variable $u$.
Depending on the physical origin of $u$, the above hamiltonian can be
used to describe various disordered elastic systems.
For example,  the field ${\bf u}$ could denote the displacements from
the
equilibrium position  for vortex lattices or could be a phase variable
in the case
of charge density waves.
$c$ is the  non-dispersive elastic modulus  and $V(x)$ is the
random potential  describing  the point-like impurities which couples
to the density $\rho$.
The  random potential is gaussian correlated
$\langle V(x) V(x^{\prime}) \rangle = W \delta(x-x^{\prime})$.
We use the decomposition of the density  in terms of harmonics of the
lattice\cite{giamarchi_vortex_short,giamarchi_vortex_long}
\begin{equation} \label{eq:deco}
\rho(x) \simeq \rho_0 (1 - \partial _{\alpha} u_{\alpha} +
\sum_{{ K}\neq { 0}} e^{i K(x - u)})
\end{equation}
where $K$ are the vectors of the reciprocal lattice and 
$\rho_0$ is the average density.
Using (\ref{eq:deco})
and the replica trick to average over
disorder we obtain the following replicated hamiltonian
\begin{equation} \label{hrep}
H_{\text{eff}}= \int d^d x  \sum_a \frac{c}2 (\partial u_\alpha^a)^2
-\sum_{a,b,{ K} \neq 0} {{\rho_0^2 W} \over {2T}}
\cos(K \cdot (u^a (x) - u^b (x)))
\end{equation}
Here $T$ is the temperature, and irrelevant terms above the 
lower critical dimension have been discarded.
To obtain the effects of disorder on all length scales  one needs to
study  the theory retaining all the
 cosine terms in the replicated Hamiltonian (\ref{hrep}).
However,  since we are mainly interested in
the asymptotic physics   which is governed  by the fundamental ${ K}$
vector ${ K}_0$, we retain only the lowest harmonic ${ K}={
K}_0$
in (\ref{hrep}).
We now study the hamiltonian (\ref{hrep}) using both the GVM and the
FRG.

\subsection{Gaussian Variational Method}

To solve (\ref{hrep}), we follow the lines of
Ref.~\onlinecite{giamarchi_vortex_short,giamarchi_vortex_long},
and introduce the variational Hamiltonian
\begin{equation}
H_0 = \frac12 \int \frac{d^d q}{(2\pi)^d} 
\sum_{ab} u^a_{\alpha}G^{-1 ab}_{\alpha
\beta}
u^b_{\beta}
\end{equation}
The indices $a,b$ denote the replicas and $\alpha,\beta$ denote the
components of the
field. Here $G$ is the variational Green's function matrix which is
parametrised as
\begin{equation}
G^{-1 }_{ab,\alpha \beta}= f_{\alpha \beta}(q) \delta^{ab} -
\sigma^{ab}_{\alpha \beta}
\end{equation}
$f_{\alpha\beta}(q)= cq^2 \delta_{\alpha \beta}$ and the $\sigma$ are
called the
self-energy parameters.
Since the disorder induced interaction
in $H_{\text{eff}}$ is  essentially local, the $\sigma$'s are all taken
to be constants.
 These parameters are determined by
 saddle point equations which  are obtained by minimising the
variational
free energy $F_{\text{var}}= F_0 + \langle H_{\text{eff}}-H_0\rangle
_{H_0}$ with respect to  $\sigma$.
The   variational equations are (see
Ref.~\onlinecite{giamarchi_vortex_long} for technical details)
\begin{eqnarray}
G_{c\alpha \beta}^{-1}&=& f_{\alpha \beta} \nonumber \\
\sigma_{\alpha\beta}^{a \neq b}(v)&=&  {{W} \over {2T}}
K_{0\alpha}K_{0\beta}
e^{[ -{1 \over 2} K_{0\alpha} K_{0\beta} B_{\alpha \beta}^{a \neq
b}(0,v)]}
\label{ve}
\end{eqnarray}
\noindent
where
$G_{c\alpha \beta}  ^{-1}(q)= \sum_b G^{-1ab}_{\alpha \beta}(q)
$ and
$B$ is defined as
\begin{equation} \label{bcor}
B_{\alpha\beta}^{ab}
= T \int {{d^dq}  \over {(2 \pi)^d}} [G_{\alpha \beta }^{aa}
+  G_{\alpha \beta }^{bb}
-2 \cos(qx)  G_{\alpha \beta }^{ab}]
\end{equation}
There are two kinds of solutions to these equations.
One is the replica symmetric and the other is the replica symmetry
broken
one.  Earlier work
\cite{giamarchi_vortex_short,giamarchi_vortex_long} has
shown that the appropriate solution has a replica symmetry broken
structure (RSB).
We thus parametrize all quantities off-diagonal in the
replica indices by hierarchical matrices
\cite{mezard_variational_replica}
\begin{eqnarray}
G^{aa}(q) &=& {\tilde G}(q) \nonumber \\
B^{aa}(x) &=& {\tilde B}(x)  \\
G^{ab}(q) &=& {G}(q,v) \nonumber
\end{eqnarray}
where $0<v<1$  and $B^{ab}(x)=B(x,v)$.
The indices $\alpha, \beta$ have been suppressed in the parametrisations
presented above.

The saddle point equations (\ref{ve}) now take the form
\begin{equation} \label{veh}
\sigma_{\alpha\beta}(v)=  {W \over {2T}} K_{0\alpha}K_{0\beta}
e^{[ -{1 \over 2} K_{0\alpha} K_{0\beta} B_{\alpha \beta}(0,v)]}
\end{equation}
where
\begin{equation}
B_{\alpha\beta}(0,v)
= 2T \int {{d^dq } \over {(2 \pi)^d}} [{\tilde G}_{\alpha \beta }(q )
-  G_{\alpha \beta }(q, v)]
\label{bcorh}
\end{equation}
Since $B(0,v)$ in  (\ref{veh}) is essentially local, it follows that
it is isotropic which results in the following simplification
i.e., $B_{\alpha \beta}(x=0)= \delta_{\alpha \beta} B(v)$ with
$B(v)$
defined below. Using this in (\ref{veh}) we find that
$\sigma_{\alpha\beta}(v)=
\delta_{\alpha\beta} \sigma(v)$.
We now search for a solution for $\sigma$  which has
full replica symmetry breaking imbedded in it. This can be  done in an
easier manner by recasting the equations in terms of
a new variable
$[\sigma]= v \sigma(v) -\int du \sigma(u)$. Plugging in the ansatz of
RSB, it can be shown that the correct solution for $[\sigma]$ has the
following  form  for $2 < d < 4$
\cite{giamarchi_vortex_short,giamarchi_vortex_long}.
\begin{eqnarray} \label{bsig}
{[}\sigma{]}(v)  &\propto & v^{\frac{2}{\theta}}, \qquad v \leq v_c \nonumber \\
{[ }\sigma {]} (v) & = & \Sigma ,\qquad  v \geq v_c
\end{eqnarray}
At the upper critical dimension there are corrections to the simple power
law.
Note that a knowledge of the exponent $\theta$, the break-point
$v_c$ and $\Sigma$  is sufficient to fix the full functional form of the
original $\sigma(v)$.
As a first step in evaluating these quantities  one  uses the
the inversion rules for  hierarchical matrices
\cite{mezard_variational_replica} to obtain
\begin{eqnarray}
B(0,v) &=& B(0,v_c) \nonumber \\
         && + \int_v^{v_c} du \int {{d^d q } \over {(2\pi)^d}}
{{2T \sigma^{\prime}(u)} \over {[G_{c}^{-1} (q) + [\sigma](u)}]^2}
\label{bcorb}
\end{eqnarray}
where
\begin{equation}
B(0,v_c)=
 \int {{d^d q } \over {(2\pi)^d}}
{{2T } \over {G_{c}^{-1} (q) + \Sigma}}
\label{bvc}
\end{equation}
\noindent
Therefore, using Eqs.(\ref{bsig},\ref{bcorb},\ref{bvc}) in  (\ref{ve})
one can obtain a self-consistent solution for $\sigma$.
Adapting the  general framework presented above to the case of
a   single component
model  where $G_c^{-1}= cq^2$  and using (\ref{bcorb})
we  find that
 the equation determining  $[\sigma]$     is
\begin{equation}
\sigma(v) \int {{d^d q} \over {(2 \pi)^d}} {{ { K}_0^2 T} \over
 {[G_c(q)^{-1} + [\sigma](v)]^2}} = 1
\label{scsig}
\end{equation}
\noindent
Solving for $[\sigma]$, we find
\begin{equation} \label{ss}
[\sigma](v)= {{Av} \over { \log^2 {{Av} \over {c{\Lambda}^2}}}}
\end{equation}
where $\Lambda$ is the ultra-violet momentum cut-off
and $A_{}=  K_0^2 T S_d / {2 c^{d \over 2}}$ where $S_d$ is the
value of the angular integration in $d$ dimensions.
For $d < 4$,  $[\sigma]$ is a \emph{pure} power law
\cite{giamarchi_vortex_short,giamarchi_vortex_long}
with the exponent $\theta=d-2$ as shown
in (\ref{bsig}). The logarithmic
corrections are specific to $d=4$ (or rather $d=D_c$ as will be
seen later) and have important
consequences for the physical properties such as the correlation
functions.
The solution (\ref{ss}) for $[\sigma]$ is
valid upto some $v=v_c$ above which
$[\sigma]$ is a constant. This constant value is again fixed by
(\ref{veh}).

Using (\ref{bcor}) and (\ref{veh}) we can now calculate the correlation
functions. The displacement correlation function is given by
\begin{eqnarray}
{\tilde B}(x)&=& \overline{\langle (u(x) - u(0))^2 \rangle}  \\
&=& 2T \int {{d^d q } \over {(2\pi)^d}}
 (1 -\cos(qx )) {\tilde G}
\end{eqnarray}
with
\begin{equation}
{\tilde G}(q) = G_c(q) [ 1+ \int_0 ^1 {{dv} \over {v^2}}
{{[\sigma](v)}
\over {[G_c^{-1} + [\sigma](v)]}}]
\end{equation}
\noindent
Since the long-wavelength behaviour of $\tilde B$  is
governed by the small $v$ behaviour of $[\sigma]$,
the  asymptotic displacement correlation function is
\begin{equation} \label{cor1}
{\tilde B}(x) = { 2 \over {K_0^2}} \log \log (\Lambda x)
\end{equation}
\noindent
This growth of the displacements is indeed  as expected
slower than the logarithmic
growth seen below four dimensions
\begin{equation}
{\tilde B}(x)=\frac{2m}{K_0^2} A_d
\log|x|
\end{equation}
with $A_d=4-d$ and $m$ is the number of components of $u$.

One can also extract \cite{giamarchi_vortex_long}
the  Larkin length $R_c$  given by $R_c=\sqrt{\frac c{\Sigma}}$
from the variational
solution  presented above. At zero temperature we obtain
\begin{equation}
R_c =  {a \over {2\pi}} \exp {{2c^2 \pi^2} \over {{ K}_0^4  W}}
\label{rcv}
\end{equation}
\noindent
where $a$ is the lattice spacing.
The exponential growth, is characteristic of the upper critical
dimension and is an extrapolation of the power law form
\begin{equation}
R_c = (c^2/W K_0^4 c_d)^{1/(4-d)}
\end{equation}
occurring for $d<4$.
The result (\ref{rcv}) given by the variational method coincides 
with the naive dimensional result or the more accurate result for 
$R_c$ obtained using the FRG procedure
\cite{bucheli_frg_rc,wagner_thesis}. 
Thus, as was already the case
\cite{giamarchi_vortex_short,giamarchi_vortex_long} for $d<4$,
the variational method gives the correct disorder 
dependence of the Larkin length.

\subsection {Functional Renormalisation Group}

We now compute the correlation functions using a functional
renormalization group procedure. The results of the GVM and the FRG
agree very well for $d<4$
\cite{giamarchi_vortex_short,giamarchi_vortex_long}.
As  is usual with renormalization procedures, we can expect the FRG to
work extremely well at the upper critical dimension. This provides
another test  to verify
whether the equivalence between the FRG and GVM results
which exists in $d < 4$  persists in $d=4$. For simplicity,
here we restrict  to a single component model.

We denote by
$\Delta(u^a (x) -u^b(x))$ the nonlinear term, proportional
to the disorder strength
in the hamiltonian (\ref{hrep}).
 For clarity we set
$K_0=2 \pi$. The function $\Delta(z)$
is initially $\sim \cos(2 \pi z)$ and the idea  is to obtain flow
equations for this function whose
fixed point solutions will  allow one obtain the correlation
functions.
In dimensions $d >4$, the gaussian fixed point $\Delta(z)=0$
is perturbatively stable, and thus the cosine potential $\Delta$ can be expanded around $z=0$.
However, for $d\leq 4$,
we need to retain all the terms in the expansion because they are
all marginal. Hence,  to study the flow of
all these terms one uses the FRG as opposed to the usual RG.
To obtain the flow equations of $\Delta$ and $T$, we use the rescaling
$x \to e^l x$ and $u \to e^{\zeta l} u$. Since the potential $\Delta$
is periodic in its argument, $u$ cannot be rescaled, implying
$\zeta=0$.
Using a momentum-shell RG \cite{fisher_functional_rg},
we find that the system is described by
a zero temperature fixed point. The corresponding flow equations for
$\Delta$ and $T$  for $d \leq 4$ is
\begin{eqnarray}
{{dT} \over {dl}}&=& (2-d) T\nonumber \\
{{d\Delta} \over {dl}}&=& \epsilon \Delta + {1 \over 2}
(\Delta^{\prime \prime})^2
- \Delta^{\prime \prime}
 \Delta^{\prime \prime}(0)
\label{rge}
\end{eqnarray}
\noindent
 where $\epsilon = 4-d$ and corrections to these equations are
of higher order in $\epsilon$.
A factor $1/S_d=2^{d-1} \pi^{d/2} \Gamma[d/2]$ has been absorbed in
$\Delta$.
 From the flow equation for $T$ we see that temperature is an irrelevant
variable for all $d>2$.
The  periodic fixed point solution to this highly non-linear
differential equation for $\Delta$,  for finite $\epsilon$
is \cite{giamarchi_vortex_short,giamarchi_vortex_long}
\begin{equation} \label{fdp}
\Delta^*={\epsilon \over {72}} ({1 \over {36}} - z^2(1-z)^2)
\end{equation}
for $z$ in the  primary interval $[0,1]$ and is extendable to other
values of
$z$ by periodicity.
The solution   in (\ref{fdp}) is  non-analytic at $z=0$ and this
non-analyticity is \emph{generic} to disordered fixed points of the kind
studied here.

As anticipated, (\ref{fdp}) shows that at the upper critical dimension
the fixed point is $\Delta(z)=0$.  Note that the RG equation in this
case ($d=4$) gives  the exact large scale behaviour while in
$d=3$ one must rely on the $O(\epsilon)$ expansion.
To find the scale dependence of
the function $\Delta_l(z)$ in $d=4$,  we use the  rescaling
\begin{equation} \label{rd}
\Delta _l^* (z) = {y \over l }
\end{equation}
Substituting this in (\ref{rge}),
the corresponding fixed point equation satisfied by  $y$ is
\begin{equation}
0 = y + (\frac{1}{2} y^{\prime \prime})^2 - y^{\prime \prime} y^{\prime \prime}(0)
\label{ey}
\end{equation}
Since this equation has a structure identical to (\ref{rge})
with $\epsilon=1$ we can use the same solution to obtain
\begin{equation}
y={1 \over {72}} ({1 \over {36}} - z^2(1-z)^2)
\label{fps}
\end{equation}
\noindent
 in the interval $[0,1]$.
Therefore, (\ref{fps}) together with the rescaling  (\ref{rd})
for $\Delta^*$
 specifies  the fixed point solution in $d=4$.
This solution also exhibits the afore-mentioned non-analytic behaviour
at the origin.
Another interesting feature of the solution to (\ref{rge}) in $d=4$ is
that the explicit $l$ dependence of the solution shows us how the
fixed point solution is approached as one scales the system.
Using the above  results, we can now calculate the displacement
correlation
function
$B(q)= \langle u^a(q) u^a(-q) \rangle$ which
satisfies the following flow equation
\begin{equation}
B(q)= e^{4l} B(qe^l,T,\Delta_l)= ({1 \over {qa}})^4
B(a^{-1},T=0,\Delta^*)
\label{cor2}
\end{equation}
\noindent
In (\ref{cor2}), we have used $e^l = (qa)^{-1}$
where $a$ is the lattice spacing which provides the ultra-violet
cut-off.
For long wavelength correlations, one can expand (\ref{cor2})
perturbatively in $\Delta^*$
 to finally obtain
\begin{equation}
B(q)= \left(\frac1{q}\right)^4 (-\Delta^{*\prime \prime}(0))
\end{equation}
Using (\ref{fps}) and restoring the factor $S_4=8 \pi^2$ this leads to
\begin{equation}
B(q)=  {{8\pi^2} \over {36}}   {1 \over {q^4 \log (aq)}}
\label{cq}
\end{equation}
\noindent
The fourier transform of (\ref{cq}) results in the following
 real space displacement correlation function
\begin{equation}
{\tilde B}(x)= \overline{\langle (u(x) - u(0))^2\rangle} =
 { {8 \pi ^2} \over {36 K_0^2}} \log \log \Lambda x
\label{cor3}
\end{equation}
\noindent
where we have restored $K_0$.
We find that both the GVM and FRG methods (\ref{cor1},\ref{cor3}) yield
the same form for
\begin{equation}
{\tilde B}(x) \simeq { b \over {{ K}_0^2}} \log \log[\Lambda x]
\end{equation}
The prefactor $b$ which also
determines the exponent in the  translation correlation
function is
$b_{var}= 2$ and $b_{frg}=2.19  $. This discrepancy  in the value of
$b$ is the same as the
one that exists at dimensions $d < 4$.
There such a difference was attributed to the fact that the variational
method  underestimates the effect of fluctuations and one expects the
the same argument to hold  in $d=4$ as well.

\section{Physical system with non-local elastic constants}
\label{sec:disp}
\subsection{Single Component Systems}

Let us now turn to a realistic system with long range interactions.
The simplest system is   a
CDW system with long range Coulomb interactions. In this case the
field
$u$ is a simple scalar field describing the phase.  Due to the long range nature of the
Coulomb interaction the elastic propagator becomes \cite{chitra_wigner_hall}
\begin{equation}
G^{-1}_c(q) = c q^2 + d q
\end{equation}
instead of the simple $q^2$  propagator
corresponding to local elasticity.
By writing $d q = \alpha(q) q^2$ with $\alpha =\frac{d}{q}$, we can
view the Coulomb contribution  as being
equivalent to a non-local elastic modulus.
When $d$ is large 
this shifts the ``physical'' upper critical dimension from $D_c=4$
to $D_c=2$. The results of the previous section are
thus
directly applicable. We find that
$[\sigma]$ has the same form as that given by (\ref{ss})
with $A= {{2 \pi d^2} \over {K_0 ^2 T}}$.
 Rather surprisingly, the
asymptotic displacement correlation function is found to grow in precisely the
same manner as the correlations in $d=4$
\begin{equation} \label{corc}
{\tilde B}(x) = { 2 \over {K_0^2}} \log \log (\Lambda x)
\label{coul}
\end{equation}
\noindent
The FRG can be used in the present case with the replacements
$\epsilon=2-d$ and $S_4 \to S_2$ in (\ref{cq}). This results in the same
correlation as in (\ref{coul}) with the prefactor $2.19$ instead of $2$. 
Note however, that in systems such as Wigner crystals
where transverse displacements exist, these transverse
displacements are insensitive to the long range part of the Coulomb
interaction and the upper critical dimension remains $D_c=4$~
\cite{chitra_wigner_hall,giamarchi_columnar_variat}, leading to a different result
for the correlations in $d=2$.

\subsection{Vortex Lattice}

\begin{figure}
\centerline{\epsfig{file=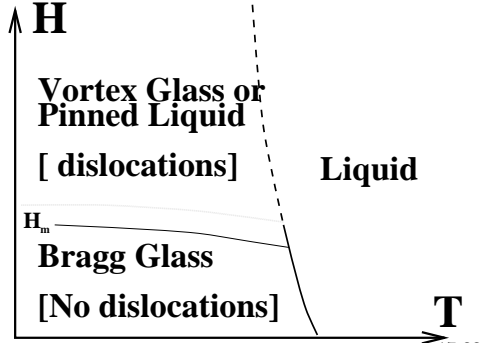,angle=-90,width=6cm}}
\caption{\label{fig:diag}
Schematic phase diagram of the $3$-d vortex lattice in the
$H-T$ plane \protect{\cite{giamarchi_vortex_long,giamarchi_diagphas_prb}}.
The stability region of the Bragg glass phase in the magnetic
field $H$, temperature $T$ plane is shown for moderate fields (the
very low field region is not shown). 
The thick line is expected to be first order, whereas the dotted line 
should be either second order or a crossover. Upon increasing
disorder the field induced melting occurs for lower fields as
indicated by the thin solid line.}
\end{figure}

We focus in this section on another physical realization of elastic
systems at the upper critical dimension,  namely, the
vortex lattice in  $d=3$.
The physical properties of such a system with non-dispersive elastic
constants  are  now well known (see e.g.
Ref.~\onlinecite{giamarchi_book_young} and references therein)
and lead to the
phase diagram of Figure~\ref{fig:diag}.
The Bragg glass occurs for
weak disorder where collective pinning  works. The system is dislocation
free and displacements are logarithmic.
At high fields the disorder is stronger and dislocations
are present.
We study here the effect of non-local elastic constants in the
Bragg glass phase. For a realistic form of the dispersion
of these moduli i.e., momentum dependence, the system  could
effectively  behave like a $d=4$ system at intermediate lengthscales.
This scenario is realised in high-$T_c$ vortex lattices.

\subsection{Model}

The prototype elastic hamiltonian for $d=3$
vortex lattices formed in
type II superconductors  with point-like disorder  is
\begin{eqnarray}
H=& {1 \over 2}&\int \frac{d^2 q_{\perp} dq_z}{(2 \pi)^3}
 [(c_{11} -c_{66}) (q_\alpha
u_\alpha)^2
    + c_{66} (q_\alpha u_\beta)^2 \\ \nonumber
& + & c_{44}(q_z u_\alpha )^2]   + \int d^2 r dz V(r,z) \rho(r,z)
\label{hvl}
\end{eqnarray}
\noindent
Here ${\bf u}$ are the
displacement vectors and $\alpha,\beta$ represent the $x,y$  components.
In (\ref{hvl}), $c_{11}$, $c_{66}$
and  $c_{44}$  are the compression, shear and  tilt moduli respectively.
These elastic constants  are in general momentum
dependent
 and typically depend on the detailed properties of
the superconductors. They  have been  evaluated  within a
Ginzburg Landau theory and are given by the following expressions
\cite{blatter_vortex_review}
\begin{eqnarray} \label{nloc}
&& c_{66}= \frac{\Phi_0 B}{(8 \pi \lambda)^2}  \nonumber \\
&& c_{44}= \frac{B^2}{4 \pi} \frac{1}{1 + \lambda^2/\epsilon^2
{q_\perp}^2 +
 \lambda^2 q_z^2} + c^s_{44} \\
&& c_{11}= \frac{B^2}{4 \pi} \frac{1 + \lambda^2/\epsilon^2 q^2}{
(1 + \lambda^2 q^2) (1 + \lambda^2/\epsilon^2 {q_\perp}^2 + \lambda^2 q_z^2)}
 \nonumber
\end{eqnarray}

Here $\Phi_0$ is the flux quantum,
$B$ is the  magnetic field along the c axis,
$\epsilon = \sqrt{M_{\perp}/M_z} \ll 1 $ is the anisotropy parameter,
 $\lambda$ is  the
London  penetration depth in the $ab$ plane and $q_{\perp}$ is the in-plane
momentum.
$c^s_{44}$ is the single vortex contribution which is negligible
for $q_{\perp} < 1/a$.
 Note that though the shear modulus is always
non-dispersive, $c_{11}$  and
$c_{44}$  are both dispersive, in other words, non-local. The origin
of this dispersion
is known to lie in the long ranged Yukawa like interaction between
vortex segments. Earlier work
\cite{larkin_70,larkin_ovchinnikov_pinning,giamarchi_vortex_short,%
giamarchi_vortex_long,nelson_columnar_long}
had focussed on regimes of the field where the dispersion
in these constants was negligible and one recovered a completely local
hamiltonian.
A logarithmic growth of the  asymptotic transverse and longitudinal
 displacements was found in this case.
This growth was also found to be isotropic despite   the   obvious
anisotropy in the
elastic hamiltonian.
 Apart from this logarithmic regime, a  random manifold regime
 with a power law growth of displacements
 for intermediate
length scales
 where all the cosine terms were
relevant
 was also found.
However, in addition to these regimes, there exists another  regime
where the
dispersive nature of the elastic constants  discussed above becomes
crucial.
This is the region where
  $c_{11}$ and $c_{66}$ are both non-dispersive
with $c_{11} \gg c_{66}$ and  $c_{44}$ is dispersive
(or $c_{11}$  is dispersive but much larger than
$c_{66}$ in which case we can set $c_{11}^{-1}=0$ in the
following). The approximation to $c_{44}$ in this regime is given by
\begin{equation} \label{appr}
c_{44}= {{B^2 \epsilon^2} \over {4 \pi \lambda^2 q_{\perp}^2}}
\end{equation}
This $q-$ dependent elastic modulus changes  the nature of
the effective long wave-length theory.  In
Ref.\onlinecite{giamarchi_vortex_short,giamarchi_vortex_long}, it
was argued that such a non-local regime might manifest itself in cases
where the translational correlation length $R_a$ was smaller than
$\frac\lambda{\epsilon}$. The effect of such non-local elasticity
on vortex lattices was reexamined recently 
\cite{bucheli_frg_rc,wagner_thesis}, and the disorder
dependence of the Larkin length was computed 
using a second order FRG calculation. We focus here 
on the calculation of the translational order parameter.
As will be shown below, correlation
functions in this regime behave very differently from  that of the
logarithmic and random manifold regimes.

The disorder independent part of the
Hamiltonian when rewritten in terms of the
transverse and longitudinal components  in Fourier space is
\begin{equation}
H_{0}= \int \frac{d^2 q_{\perp} dq_{z}}{(2 \pi)^3}   
(c_{66}q_{\perp}^2+c_{44}q_z ^2) u_T^2
+   (c_{11}q_{\perp}^2+c_{44}q_z ^2) u_L^2
\end{equation}
\noindent
Using the formalism presented in Sec. II
 and  the following connected Green's functions
\begin{eqnarray}
G_{cT}^{-1}&=& c_{44}q_z^2 +c_{66}q_{\perp}^2 \nonumber \\
G_{cL}^{-1}&= &c_{44}q_z^2 +c_{11}q_{\perp}^2
\label{gconn}
\end{eqnarray}
\noindent
 we find that
$[\sigma]$ in the present case is determined by the following
transcendental
equation (refer to the Appendix for details)
\begin{equation}
[\sigma] \log^2 { {\Lambda^2} \over {[\sigma]}} [c_{11}^{-3/2} +
c_{66}^{-3/2}]
=  {{64 \pi \sqrt{\Gamma}} \over {{ K}_0^2 T}} v
\label{vsig}
\end{equation}
\noindent
where $\Gamma = B^2 \epsilon^2/(4 \pi \lambda^2)$
and $\Lambda = {{2\pi} \over a}$ is the lattice cut-off.
Solving (\ref{vsig}), we find that $[\sigma]$ has the same form as that
given in (\ref{ss}), with the constant $A$ now given by
\begin{equation}
A_v= {{64 \pi} \over {{ K}^2_0 T}} \sqrt{\Gamma} [{1 \over
{c_{11}^{3 \over 2}}}
+ {1 \over {c_{66}^{3 \over 2}}}]^{-1}
\end{equation}
Using the result for $[\sigma]$  the displacement correlation
functions are  given by
\begin{eqnarray}
{\tilde B}_{L,T}(r,0)=  2T \int {{d^2 q_\perp dq_z} \over {(2\pi)^3}}
[&&\cos^2(\theta) (1 -\cos(q_\perp r \cos(\theta)))
    {\tilde G}_{L,T} \\
&& + (1- \cos^2 (\theta))
 (1 -\cos(q_\perp r \cos(\theta))) {\tilde G}_{T,L}]
\label{nlc}
\end{eqnarray}
\noindent
Here
\begin{equation}
{\tilde G}_{L,T}(q_\perp,q_z)= \int ^1 _0 {{dv} \over {v^2}} {{[\sigma](v)}
\over {G^{-1}_{cL,T} + [\sigma]}}   {G_{cL,T}}
\label{nlg}
\end{equation}

Performing  a whole lot of cumbersome integrations (see Appendix), we
finally obtain the following
 leading terms in the displacement correlation functions:
\begin{equation}
{\tilde B}_{L,R} (r,0)  =
{\tilde B} (r,0)= {2 \over {{ K}_0^2}} \log \log {\Lambda^2 r^2}
\label{vcor}
\end{equation}
\noindent
Note that this behaviour of the correlations is exactly analogous to the
ones seen earlier in $d=4$  (cf. (\ref{cor1}),(\ref{cor3})) and
$d=2$ proving again 
that non-local elastic constants of the type considered here
increase the effective dimensionality of the system.
Note that despite the  strong anisotropy in the  hamiltonian,
isotropy is restored in the the long wavelength  correlation functions
in this dispersive regime too. Another interesting
point is that these correlations are
universal in that they do not depend on the value of  any parameter in
the theory.
The above result  can  now be used to calculate the translational order
parameter. Within the Gaussian approximation it is thus given by:
\begin{eqnarray}
O_T &=& \langle \exp {i K_0.({\bf u}(r)- {\bf u}(0))} \rangle =
\exp-{\frac{K_0^2}{2}} \langle ({\bf u}(r,0)-{\bf u}(0,0))^2\rangle \\
&\simeq&
\frac{1}{(\ln r)}
\label{vtr}
\end{eqnarray}
Note that the decay of $O_T$ in the present case is much slower than the
$\frac1{r}$ decay seen in the case where the
elastic constants are local. This implies that  the system will look
more ordered than it does in the other regimes.

\subsection{Physical Discussion}

We now  estimate the regime of validity of the
previous calculation i.e., the range in which the dispersive
approximation to $c_{44}$ and hence the $\log\log$ growth of
displacements  are  valid (for a very complete description
of the possible regimes in anisotropic superconductors 
see Ref.~\onlinecite{wagner_thesis}).
Re-analysing  the expressions (\ref{nloc}) for the
elastic constants
we see that the approximation (\ref{appr}) to $c_{44}$
is valid provided one is working at length scales $L_\perp$ and $L_z$ which
satisfy the following bounds
\begin{eqnarray}  \label{regime}
&& L_{\perp} \ll \lambda/\epsilon \\
&& L_{\perp} \ll L_z/\epsilon
\end{eqnarray}
It is also interesting to check when the above dispersive regime
includes both the  Larkin lengths $R_\perp$ and $R_z$.
The Larkin lengths are
determined by the balance between the elastic energy and the
disorder which in $d=3$ takes the form
\begin{equation}
c_{66} r_f^2 R_z^{-2} \sim c_{44} r_f^2 R_{\perp}^{-2}
\sim r_f R_{\perp} R_z^{1/2} W^{1/2}
\end{equation}
\noindent
Here $r_f$ is the length over which disorder is correlated.
We immediately see that $R_z/R_\perp = \sqrt{c_{44}/c_{66}}$
and  using (\ref{appr}) we  thus obtain
\begin{equation}
R_z = \epsilon R_{\perp}^2 /a
\end{equation}
\noindent
This in conjunction with (\ref{regime}) yields the following  bounds
on $R_\perp$
\begin{eqnarray}
a \ll R_\perp \ll \lambda/\epsilon
\end{eqnarray}
\noindent
The upper limit corresponds to $R_z = \lambda^2/(\epsilon a)$ and the
lower limit $R_z = \epsilon a$. The latter may be smaller than
the distance $d$ between layers resulting in  complicated
dimensional crossovers. To stay away from possible
dimensional crossover we will restrict $R_\perp$ such that
\begin{eqnarray}
\sqrt{a d/\epsilon} \ll R_\perp \ll \lambda/\epsilon
\end{eqnarray}
\noindent
which corresponds to $d < R_z < \lambda^2/(\epsilon a)$. Note that
the above regime becomes non-existent  above the ``decoupling crossover
field''
$B_{cr} = \Phi_0 \epsilon^2/d^2$.
\begin{figure}
\centerline{\epsfig{file=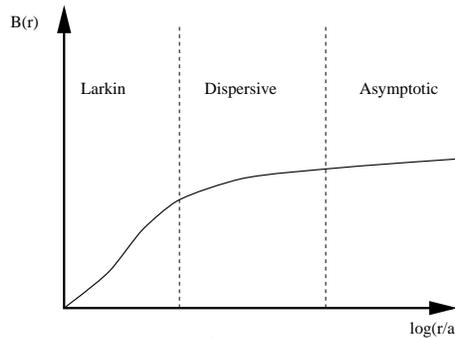,angle=270,width=6cm}}
\caption{\label{fig2} 
The growth of the displacements $\tilde B(r)$ as a function
of $\log(r/a)$ in various regimes.}
\end{figure}
We therefore see that  the result of (\ref{vcor})  can be used
at all lengths scales within the bounds prescribed by (\ref{regime}).
For asymptotically large scales, the elastic constants become
non-dispersive and
one expects a crossover to the simple logarithmic growth
of displacements discussed in
Ref.\onlinecite{giamarchi_vortex_short,giamarchi_vortex_long}.
These results are
summarised in Fig.2 .

\section {Conclusions} \label{sec:conc}

We have studied the long wavelength correlations of elastic systems at
their upper critical dimension in the presence of weak pointlike
disorder.
The cases analysed include the simple elastic system at $d=4$,
single component systems with $\frac1{r}$ interactions and vortex lattices
with non-local elastic constants.
The systems were studied using the GVM and the equivalence between the
GVM and the FRG methods was explicitly demonstrated in the simpler
cases of  $d=4$  and $d=2$ and we
 expect such a correspondence to hold in the other cases as well.
In all the cases studied we observe the striking fact that the
growth of asymptotic
displacements is generic in that it is always $\log \log(x)$.
This was found to be true even in the case of the
anisotropic
vortex lattice, where  isotropy was
recovered in the asymptotic correlation functions.
In addition,
the prefactor of this $\log \log$ term (which determines the decay of
the translational order within the GVM) is independent of the parameters
of the models considered.
These systems are again  Bragg glasses  and the peaks in the diffraction
will be  more pronounced  because the systems are more ordered than
their other disordered counterparts which typically show a
logarithmic growth of displacements.
This  feature should be
verifiable using decoration and diffraction experiments.

\section{Appendix}

We first write down the explicit form of the equation which determines 
$[\sigma]$ in the case of the vortex
lattice described in Sec.\ref{sec:disp}. 
This is a generalisation of (\ref{scsig}) 
to the two component case and it reads 
\begin{equation}
\sigma(v)
 {{ { K}_0^2 T} \over 2}
\int {{d^d q} \over {(2 \pi)^d}} [{1 \over
 {(G_{cT}(q)^{-1} + [\sigma](v))^2}} +
{1 \over
 {(G_{cL}(q)^{-1} + [\sigma](v))^2}}] = 1
\end{equation}
\noindent
Using the expressions for $G_{cT,L}$ given in (\ref{gconn}) and integrating
over $q_\perp$ and $q_z$ we arrive at (\ref{vsig}).

 Next we present the details of the calculation of the displacement
correlation
function discussed in Sec.V .
Performing the angular integrations in (\ref{nlc})we get
\begin{eqnarray}
{\tilde B}_L= {{2T} \over {(2\pi)^2}} \int q_\perp dq_\perp dq_z&& [({1 \over 2} - 
J_0(q_\perp r) +
{{J_1 (q_\perp r)}  \over {q_\perp r}}) {\tilde G}_L\\
&& +  ({1 \over 2}  -
{{J_1 (q_\perp r)}  \over {q_\perp r}})] {\tilde G}_T
\label{a1}
\end{eqnarray}
\noindent
$J_0$ and $J_1$ are the usual Bessel functions and
the expressions for $\tilde G_{L,T}$ are given in (\ref{nlg}).
(\ref{a1})    is rewritten as
\begin{equation}
{\tilde B}_L = {{2T} \over {(2\pi)^2}} \int {{dv} \over {v^2}} ( I_1
+I_2)
\label{a2}
\end{equation}
\noindent
where
\begin{equation}
I_1= \int  q_\perp dq_\perp dq_z [{{q_\perp^2} \over {\Gamma q_z^2
+ c_{11}q_\perp^4}}][
{{q_\perp^2} \over { \Gamma q_z^2 +c_{11}q_\perp^4 +[\sigma] q_\perp^2}}]({1 \over 2} -J_0
+{{J_1} \over {q_\perp r}})
\end{equation}
and
\begin{equation}
I_2= \int  q_\perp dq_\perp dq_z [{{q_\perp^2} \over {\Gamma q_z^22 +
c_{66}q_\perp^4}}][
{{q_\perp^2} \over { \Gamma q_z^2 +c_{66}q_\perp^4 +[\sigma] q_\perp^2}}]({1 \over 2} -
{{J_1} \over {q_\perp r}})
\end{equation}
Integrating over $q_z$ we get
\begin{equation}
I_1= {\pi \over {2 \sqrt{\Gamma c_{11}}}} \int q_\perp dq_\perp [ {{\sqrt{c_{11}
q_\perp^2 +
[\sigma]} -{\sqrt{c_{11}}}q_\perp} \over {\sqrt{c_{11}q_\perp^2 +[\sigma]}}}]
({1 \over 2} -J_0 +{{J_1} \over {q_\perp r}})
\end{equation}
\noindent
$I_2$ has the same form with $c_{11} \leftrightarrow c_{66}$ and the
Bessel
functions appropriately changed.
Integrating with respect to $q_\perp$ we now get
\begin{eqnarray}
I_1=&& {\pi \over {2\sqrt{\Gamma c_{11}}}} [ {1 \over{r^2}} +
{{[\sigma]} \over {4c_
{11}}} \log {{2 \Lambda} {\sqrt{{c_{11}} \over {[\sigma]}}}} \\
&&- {1 \over {2r}}
{\sqrt{{[\sigma]}
\over {c_{11}}}} (I_0(\alpha) K_1 (\alpha) - I_1(\alpha) K_0(\alpha))\\
&&- {{[\sigma]} \over {4c_{11}}} (I_0(\alpha)K_0(\alpha)- I_1(\alpha)
K_1(\alpha))]
\end{eqnarray}
and
\begin{eqnarray}
I_2=&& {\pi \over {2\sqrt{\Gamma c_{66}}}} [ -{1 \over{r^2}} +
{{[\sigma]} \over {4c_{66
}}} \log {{2 \Lambda} {\sqrt{{c_{66}} \over {[\sigma]}}}} \\
&&+ {1 \over {2r}} {\sqrt{{[\sigma]}
\over {c_{66}}}} (I_0(\beta) K_1 (\beta) - I_1(\beta) K_0(\beta))
\end{eqnarray}
\noindent
Here  $\Lambda$  is the momentum cut-off and the $I$ and $K$ are the
modified Bessel
functions  whose arguments
 are $\alpha= {r \over 2} \sqrt{{[\sigma]} \over {c_{11}}}$ and
 $\beta= {r \over 2} \sqrt{{[\sigma]} \over {c_{66}}}$.
Finally, adding up the contributions $I_1$ and $I_2$
we get
\begin{equation}
{\tilde B}_L (r,0)= {T \over {4\pi {\sqrt{\Gamma}}}} \int {{dv} \over
{v^2}}
(I_1 +I_2)
\end{equation}
The transverse correlation function ${\tilde B}_T$ is obtained from the
above expression by interchanging $c_{11}$ and $c_{66}$.
Using the  solution for $[\sigma]$ given by (\ref{vsig})
and
\begin{equation}
{\tilde B}_L = {{A_vT}  \over {32\pi {\sqrt{\Gamma}}}} \int {{dv}
\over {v^2}}
[\log {{A_vv} \over {\Lambda^2}}]^{-2} [ {E \over 2} \log {{r^2
\Lambda^4 } \over
{A_vv}} + {\bf C} E + ...
\end{equation}
\noindent
where $E= c_{11}^{-3/2} +c_{66}^{-3/2}$.
Performing the $v$ integration gives us the result
\begin{equation}
{\tilde B}_{L,T} (r,0) \simeq {2 \over { K}^2_0} \log \log \Lambda^2
r^2
\end{equation}


\end{document}